\documentclass[twocolumn,prl,aps,superbib,tightenlines,floatfix]{revtex4}
\usepackage{amsfonts,amsmath,amssymb,amsthm,bm}
\usepackage{graphicx}
\usepackage[dvipsnames,usenames]{color}
\begin{document}
\bibliographystyle{apsrev}
\title{Do Thermoelectric Materials in Nanojunctions Display Material Property or Junction Property?}
\author{Yu-Chang Chen}
\email{yuchangchen@mail.nctu.edu.tw}
\author{Yu-Shen Liu }
\affiliation{Department of
Electrophysics, National Chiao Tung University, 1001 Ta Hsueh Road,
Hsinchu 30010, Taiwan }
\begin{abstract}

The miniaturization of thermoelectric nanojunctions raises a fundamental question:
do the thermoelectric quantities of the bridging materials in nanojunctions
remain to display material properties or show junction properties?
In order to answer this question, we investigate the Seebeck coefficient $S$ and
the thermoelectric figure of merit $ZT$ especially in relation to the length characteristics
of the junctions from the first-principles approaches. For $S$, the metallic atomic
chains reveal strong length characteristics related to strong hybridization in
the electronic structures between the atoms and electrodes, while the insulating
molecular wires display strong material properties due to the cancelation of
exponential scalings in the DOSs. For $ZT$, the atomic wires remain to show strong
junction properties. However, the length chrematistics of the insulation molecular
wires depend on a characteristic temperature $T_{0}= \sqrt{\beta/\gamma(l)}$ around $10~$K.
When $T \ll T_{0}$, where the electron transport dominates the thermal current,
the molecular junctions remain to show material properties. When $T \gg T_{0}$,
where the phonon transport dominates the thermal current, the molecular junctions
display junction properties.

\end{abstract}
 \maketitle

Nanoscale thermoelectric devices can be considered as the new types of devices
which can be embedded into integrated chip sets to assist the stability of
devices by converting the accumulated waste heat into useful electric energy.
There has been an ever increasing interest in the thermoelectric properties
of nanojunctions \cite{Paulsson,Zheng,Wang,Galperin,Pauly,Finch,Troels,Dubi,Ke,Liu},
partially motivated by the recent experiments demonstrating
the capability of measuring the Seebeck coefficients in the molecular
junctions \cite{Reddy1,Reddy2}. As the Seebeck coefficients are relevant
not only to the magnitude but also to the slope of density of states (DOSs),
they can reveal more detailed information about the electronic structures of
the materials sandwiched in the nanojunctions beyond what the
conductance measurements can provide. The Seebeck coefficients
have been applied to explore the electronic structures of molecular junctions
using functional substitutions for the bridging molecules \cite{Liu,Reddy2}.
Theorists have proposed using gate fields and external biases as means to
modulate the Seebeck coefficients in nanojunctions \cite{Zheng,Wang,Liu}. Much
research has been devoted to the study of Seebeck
coefficients \cite{Paulsson,Zheng,Wang,Galperin,Pauly,Finch,Troels,Dubi,Ke,Liu},
but little is known about the fundamental thermoelectric properties in nanojunctions.

Indeed, the miniaturization of thermoelectric nanojunctions raises a
fundamental question: does bridging thermoelectric material in nanojunctions
show material properties or junction properties? Thermoelectric bulk crystals
usually show material properties, where the thermoelectric physical
quantities are irrelevant to the sizes and shapes of materials. In addition,
recent experiments on Seebeck coefficients in molecular junctions
also reveal strong signals of material properties. These experiments
have observed that Seebeck coefficients are insensitive to the number of
molecules in junctions and show rather weak dependence on the lengths of
the bridging molecules, which is in sharp contrast to the conductance which
shows strong exponential dependence on the lengths of
molecules \cite{Reddy1,Reddy2,Reddy3}. However, the bridging materials in
nanojunctions may have strong interactions with the contacts. From
this point of view one can say thermoelectric quantities can display
the junction characteristics. Considering the examples and the reason
quoted above, it is therefore not obvious whether the thermoelectric
quantities of the bridging materials in nanojunctions display junction
properties or material properties.

In this letter, we will show that the thermoelectric quantities in nanojunctions
unnecessarily display entire material properties or junction properties.
To demonstrate this point, this study investigates two important thermoelectric
quantities, the Seebeck coefficient ($S$) and the thermoelectric figure of
merit ($ZT$), in (metallic) aluminum atomic junctions and (insulating) molecular
junctions. It shows that metallic atomic chains reveal strong junction properties
while the insulating molecular wires partially possess the material properties,
where $S$ reveals the material property and $ZT$ displays the junction property
at temperatures larger than the characteristic temperature $T_{0}$.

To answer the question, we have developed a theory with analytical expressions
for $S$ and $ZT$ allied to a fully self-consistent first-principles calculation
in the framework of the density functional theory (DFT).
It allows us to numerically calculate $S$ and $ZT$ and subsequently investigate
these quantities analytically. We focus on the subject on whether $S$ and $ZT$
depend on the characteristics of the junctions, especially on the
dependence on length-characteristic of junctions. Before turning to the detailed
discussion, let us begin with a brief introduction on how to calculate
$S$ and $ZT$. First, we consider that the junction consists of source-drain electrodes,
with distinct chemical potentials $\mu _{L(R)}$ and temperatures $T_{L(R)}$,
as independent electron and phonon reservoirs. When an additional
infinitesimal temperature $\Delta T$ is applied across the junction, an extra
voltage $\Delta V$ is induced to compensate the electric current induced by
the temperature gradient $\Delta T$ across the junction. We then derive the
expressions for $S$ (defined as $S=\Delta V/\Delta T$) and
the electron thermal conductance (defined
as $\kappa_{el}=\Delta J_{Q}^{el}/\Delta T$), where $J_{Q}^{el}$
is the thermal current conveyed by the electrons which also carry the electric current).
\begin{equation}
S=-\frac{1}{e}\frac{\frac{K_{1}^{L}}{T_{L}}+\frac{K_{1}^{R}}{T_{R}}}{%
K_{0}^{L}+K_{0}^{R}},  \label{S}
\end{equation}
\begin{equation}
\kappa _{el}=\frac{1}{h}\sum_{i=L,R}(K_{1}^{i}eS+\frac{K_{2}^{i}}{T_{i}}),
\label{kel}
\end{equation}
where $K_{n}^{L(R)}=-\int dE\left( E-\mu _{L(R)}\right) ^{n}\frac{\partial
f_{E}^{L(R)}}{\partial E}\tau (E)$, and the transmission function
$\tau (E)=\tau ^{R}(E)=\tau ^{L}(E)$, which is a direct consequence of the
time-reversal symmetry. It has been assumed that the left and right
electrodes serve as independent electron and phonon reservoirs where
the electron population is described by the Fermi-Dirac distribution
function, $f_{E}^{L(R)}=1/\left[ \exp \left( \left( E-\mu _{L(R)}\right)
/k_{B}T_{L(R)}\right)+1\right] $, and $k_{B}$ is the Boltzmann constant.
The transmission functions are computed using the wave functions obtained
self-consistently in the DFT framework \cite{Lang,Chen1}. We should notice
that the above equations are suitable for describing $S$ and $\kappa_{el}$
in nanojunctions operated under finite external biases, where two electrodes
can have different temperatures.

In addition, the differential conductivity, typically insensitive to temperature
in cases where direct tunneling is the major transport mechanism, can be expressed as
\begin{equation}
\sigma =\frac{e}{2}\int dE\sum_{i=L,R}f_{E}^{i}(1-f_{E}^{i}) \tau
(E)/k_{B}T_{i}. \label{sigma}
\end{equation}

So far, the physical quantities ($S$, $\sigma$, and $\kappa_{el}$)
which have been discussed are related to the electron transport . It must be noted
that the heat current is conveyed by the electrons and phonons simultaneously.
The phonon thermal conductance ($\kappa_{ph}$) usually dominates the combined
thermal conductance $\kappa=\kappa_{el}+\kappa_{el}$ at large temperatures.
The complete discussion on $ZT$ shall include the essential
ingredient $\kappa_{ph}$; thus $ZT$ can be expressed as follows:
\begin{equation}
ZT=\frac{S^{2}\sigma }{\kappa _{el}+\kappa _{ph}}T,  \label{zt}
\end{equation}
where $T=(T_{L}+T_{R})/2$ is the average temperature of the
source-drain electrodes. We estimate the phonon thermal conductance
following the approaches of Patthon and Geller \cite{Patton}. It is
assumed that the nanojunction is a weak elastic link, with a given stiffness
which can be evaluated from the total energy calculations, attached to the
electrodes modeled as phonon reservoirs. The phonon thermal conductance
(defined by $k_{ph}=\Delta J_{Q}^{ph}/\Delta T$) is given by:
\begin{equation}
\kappa _{ph}=\frac{\pi K^{2}}{\hbar k_{B}}\int
dEE^{2}N_{L}(E)N_{R}(E)\sum_{i=L,R}\frac{n_{i}(E)(1+n_{i}(E))}{T_{i}^{2}},
\label{kph}
\end{equation}
where $n_{L(R)}= 1/[\exp(E/k_{B}T_{L(R)})-1]$ and
$N_{L(R)}(E)\approx C\cdot E$ is the Bose-Einstein distribution function
and the spectral density of phonon states in the left (right)
electrode, respectively. The stiffness of the bridging nanostructure
is $K=YA/l$, where $Y$ is the Young's modulus and $A$
$(l)$ is its cross-section (length).

\begin{figure}
\includegraphics[width=9cm]{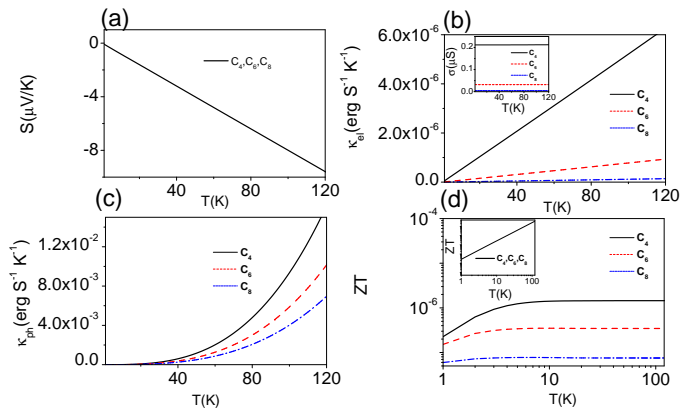}
\caption{(color online) Alkanethiol junctions
at $V_{B}=0.01$~V for C$_{4}$ [solid (black) lines], C$_{6}$ [dotted-dash (red) lines],
and C$_{8}$ [dash (blue) lines]: (a) the Seebeck coefficient $S$ vs. $T$;
(b) the electron thermal conductivity $\kappa_{el}$ and the electric
conductance $\sigma$ (inset) vs. $T$; (c) the phonon thermal conductance
$\kappa_{ph}$ vs. $T$; and (d) the log-log plot of $ZT$ vs. $T$
(inset shows the case of neglecting the thermal conductance $\kappa_{ph}=0$).}
\label{Fig1}
\end{figure}

\begin{figure}
\includegraphics[width=9cm]{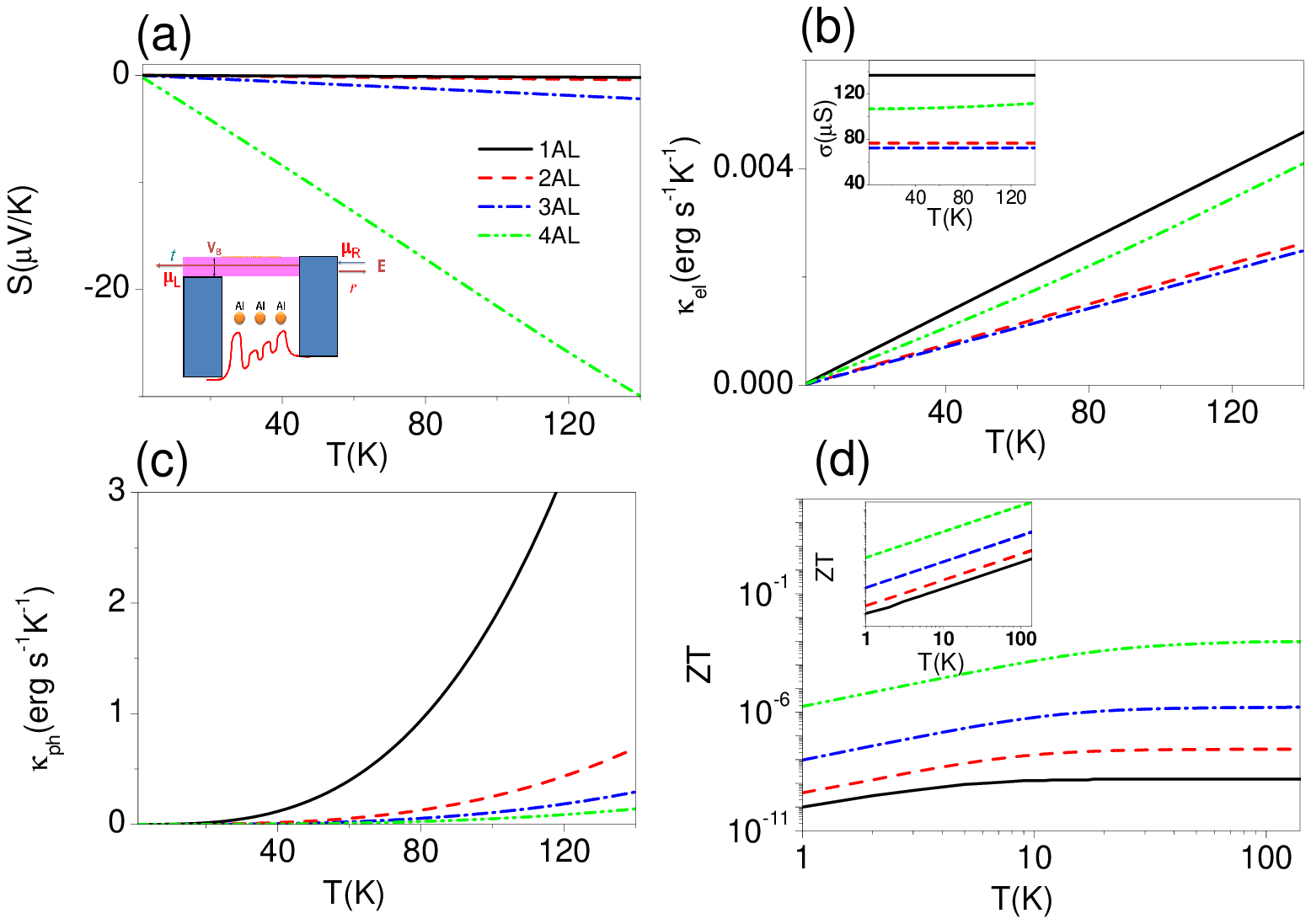}
\caption{(color online) Aluminum atomic junctions at $V_{B
}=0.01$~V for $1~Al$ [solid (black) lines], $2~Al$ [dotted-dash (red) lines],
$3~Al$ [dash (blue) lines], and $4~Al$ [dot-dot-dash (green) lines]:
(a) the Seebeck coefficient $S$ vs. $T$ and the schematic of 3-Al
atomic chain and its energy diagram (inset) where the $Al-Al$ bond
distance is about $6.3$~a.u.; (b) the electron thermal conductance
$\kappa_{el}$ and the electrical conductivity $\sigma$ (inset) vs. $T$; (c)
the phonon thermal conductivity $\kappa_{ph}$ vs. $T$; (d) the log-log
plot of $ZT$ vs. $T$ (inset shows the case of neglecting the thermal
conductance $\kappa_{ph}=0$).}
\label{Fig2}
\end{figure}

We have numerically computed $S$ and $ZT$ using Eqs.~$(\ref{S})$ to $(\ref{kph})$
allied to the transmission functions obtained self-consistently
in the DFT framework, as shown in Figs.$~\ref{Fig1}$ and $\ref{Fig2}$.
To elaborate the properties of the Seebeck coefficient and the the thermoelectric
figure of merit, we will limit our discussion to the linear response regime (i.e.,
$\mu_{L}\approx\mu_{R}=\mu$) and $T_{L}=T_{R}=T$. After expanding $S$, $\kappa_{el}$,
and $\kappa_{ph}$ in terms of the temperature $T$, we obtain the analytical expressions for
the $S$ and $ZT$.
\begin{equation}
S \approx \alpha T,
\label{S2}
\end{equation}
where $\alpha =-\pi ^{2}k_{B}^{2}\frac{\partial \tau (\mu )}{\partial E}/\left( 3e\tau (\mu )\right)$.
We have noted that the Seebeck coefficient depends on the magnitude and the
slope of the transmission function, and is linearly proportional to $T$ at
low temperatures.
\begin{equation}
ZT\approx \frac{\alpha ^{2}\sigma T^{3}}{\beta T+\gamma (l)T^{3}},
\label{ZT2}
\end{equation}
where we have expanded $\kappa_{el}$ and $\kappa_{ph}$ up to the lowest
order in temperatures as $\kappa _{el}\approx \beta T$
and $\kappa _{ph} \approx \gamma (l)T^{3}$.
The prefactor $\beta$ and $\gamma (l)$ are $\beta=2\pi ^{2}k_{B}^{2}\tau (\mu )/(3h)$ and $\gamma
(l)=8\pi ^{5}k_{B}^{4}C^{2}A^{2}Y^{2}/(15\hbar l^{2})$, respectively.
One may notice that there is a characteristic temperature
$T_{0}= \sqrt{\beta/\gamma(l)}$, which is around $10~$K for the alkanethiol molecular
junctions and is negligibly small for aluminum atomic junctions.
When $T\ll T_{0}$, the electron thermal conductance dominates
and $ZT\approx \sigma S^{2}T/k_{el}\approx \left[ \alpha ^{2}\sigma /\beta \right] T^{2}$,
which is irrelevant to the length-characteristic of the junction and is proportional
to $T^{2}$ as temperatures increase.
When $T\gg T_{0}$, the phonon thermal conductance dominates and $ZT$ tends to have a saturation
value of $ZT\approx\sigma S^{2}T/\kappa_{ph}\approx\alpha^{2}\sigma/\gamma(l)$,
which is related to the length of the junction. The above analytic expressions provide
a convenient means for analyzing the length characteristic of $S$ and $ZT$ in nanojunctions.
%%%%%%%%%%%%%%%%%%%%   Bigin GRAPHs     %%%%%%%%%%%%%%%%%%%%%%%%%%%%%%
%%%%%%%%%%%%%%%%%%%%%%%%%%%%%%%%%%%%%%%%%%%%%%%%%%%%%%%%%%%%%%%%%%%%%%
%%%%%%%%%%%%%%%%%%%%%%%%  End GRAPHs %%%%%%%%%%%%%%%%%%%%%%%%%%%%%%%%%%

Before turning to the detailed discussion on the thermoelectric properties
in nanojunctions, we must draw attention to the reason why nanojunctions
display material properties for $S$ and $ZT$. The Seebeck coefficients are
directly related to the transmission functions, which are determined by the
electronic structures of the crystal materials irrelevant to the size and
shape of the material, leading to the material properties for $S$.
Furthermore, the conductance $\sigma$
and the combined thermal conductance $\kappa=\kappa_{el}+\kappa_{ph}$ are
proportional to the contact surface and inversely proportional to the length
scale in the bulk crystal materials, which leads to the material properties
for $ZT$ due to the cancelation of the geometric
factors in the conductance $\sigma$ and the combined thermal conductance
$\kappa$. However, whether the thermoelectric materials in nanojunctions
remain to display material properties or show junction properties have
not yet been fully realized. Now, let us look closely into the
properties of $S$ and $ZT$ in the linear response regime at low temperatures
with $T=T_{L}=T_{R}$ for
the two catalogs of nanojunctions: the (insulating) alkanethiol
junctions and the (metallic) aluminum atomic junctions.

First, we examine the dependence of $S$ and $ZT$ on the lengths of the alkanethiol
junctions \cite{parametersAlatoms}. Alkanethiols [CH$_{3}$(CH$_{2}$)$_{n-1}$SH,
denoted as C$_{n}$] are a good example of reproducible junctions which can be
fabricated \cite{WangW,Tao1}. It has been well established that the non-resonant
tunneling is the main conduction mechanism in the C$_{n}$ junctions. As shown in
the inset of Fig.~\ref{Fig2}(b), the conductance is small, insensitive
to temperatures, and decreases exponentially with the length of the molecules,
as $\sigma =\sigma _{0}\exp \left( -\xi l\right)$, where
$l$ is the length of C$_{n}$ molecule and $\xi \approx 0.78$ \r{A}$^{-1}$ \cite{Wold1,Beebe,Zhao,Kaun,Ma}.
By exploiting the periodicity in the $\left(CH_{2}\right) _{2}$ group
of the C$_{n}$ chains, the wave functions of the C$_{n}$ junctions
are calculated by a simple scaling argument, leading to exponential
scaling in the transmission functions $\tau (E)$. It must be noted that
the Seebeck coefficients are proportional to $\frac{{\partial \tau (\mu )/\partial E}}{{\tau (\mu )}}$.
Consequently, the Seebeck coefficient does not show dependence on
the length [see Fig.~\ref{Fig1}(a)] of the C$_{n}$ junctions due to the
cancelation of the exponential scaling behavior in $\tau (\mu )$ for
insulating molecules. We have noted that the cancelation may not be complete
in the real experiments; thus, the Seebeck coefficients
could possibly show weak length dependence due to other effects \cite{Reddy1,Reddy2}.
In such case, the Seebeck coefficients mostly display material properties.

Let us now investigate the length-dependence of $ZT$ for the C$_{n}$ junctions.
We predict that the C$_{n}$ junctions display the material properties when $T\ll T_{0}$
and display the junction properties when $T\gg T_{0}$. The reason for this is due to
different scaling behaviors on the length characteristics of the junctions for the electron
and phonon thermal conductance and the competition between them.
When $T\ll T_{0}$, the electron transport
dominates the thermal current, and thus $ZT\approx\sigma S^{2}T/\kappa_{el}$.
As shown in the inset of Fig.~\ref{Fig1}(d), $ZT$ is independent from the
lengths of C$_{n}$ molecules and displays material properties. This result
can be explained quite naturally by the cancelation of the exponential
scaling in $\sigma$ and $\kappa_{el}$ because both the electric current
(with conductance $\sigma$) and the electron thermal current (with thermal
conductance $\kappa_{el}$) are conveyed by electron transport. When $T\gg T_{0}$,
the phonon transport dominates the thermal current and thus $ZT\approx\sigma S^{2}T/\kappa_{ph}$.
As shown in the main body of Fig.~\ref{Fig1}(d), $ZT\propto l^{2} \exp \left( -\xi l\right)$
because of $\sigma \propto \exp \left( -\xi l\right)$ [see the inset of Fig.~\ref{Fig1}(b)],
and $\kappa_{ph}\propto l^{-2}$ [see Fig.~\ref{Fig1}(c)]. In this case,
$ZT$ displays junction properties.

Finally, we will examine the dependence of $S$ and $ZT$ on the lengths
of the metallic aluminum (Al) atomic junctions \cite{parametersAlatoms}.
An Al atomic chain is an ideal testbed for studying the charge transport
at the atom-scale level [see the inset of Fig.~\ref{Fig2}(a) for a
schematic of the aluminum atomic junction] \cite{Lang1,Koba,Yang1,Cuevas}.
This study has observed that $S$ and $ZT$ depend on the geometric characteristic
of the junctions and show the junction properties in any case, which is in
sharp contrast to the C$_{n}$ junctions.
At a fixed temperature, it has been observed that $ZT$ and the magnitude of $S$
increase as the number of Al atoms increases, as shown in Figs.~\ref{Fig2}(a) and (d).
The increase of the Seebeck coefficients is due to the increase of the slopes
in the transmission functions at the Fermi levels. The reason for this may be due
to the strong hybridization between the electronic structures of atoms and the
electrodes. The negative sign of the Seebeck coefficients indicates that the
metallic Al atomic junctions are n-type [the Fermi energy is closer to the
lowest unoccupied molecular orbital (LUMO)].

In conclusion, this study has raised and answered an important fundamental question
in thermoelectric nanojunctions: do the thermoelectric materials
in nanojunctions remain to display material properties or show junction properties?
To answer this question, we have developed a theory with analytical
expressions for $S$ and $ZT$ allied to a fully self-consistent first-principles
calculation in the DFT framework. Using the insulating alkanethiol molecular junctions
and metallic aluminum junction as examples, this study concludes that the thermoelectric
quantities in nanojunctions do not necessarily display material properties or junction
properties. The metallic atomic chains reveal strong length characteristics related to the
strong hybridization in the electronic structures between the atoms and electrodes, while
the insulating molecular wires display strong material properties due to the cancelation of
exponential scalings in the DOSs. For $S$, the metallic atomic chains reveal strong length
characteristics related to the strong interactions between the electronic structure of
atoms and the electrodes, while the insulating molecular wires show independence from
the lengths of molecules, thus displaying strong material properties due to the cancelation
of exponential scalings in the density of states. It may be worth pointing out the cancelation
may not be complete in real experiments due to other effects; the important point is,
the strong exponential scaling behavior with the lengths of the junctions should be canceled.
For $ZT$, the atomic wires remain to show strong junction properties. However, the length
characteristics of the insulation molecular wires depends on a characteristic
temperature $T_{0}= \sqrt{\beta/\gamma(l)}$ around $10~$K.
When $T \ll T_{0}$, where the electron transport dominates the thermal current,
the molecular junctions remain to show material properties. When $T \gg T_{0}$,
where the phonon transport dominates the thermal current, the molecular junctions
display junction properties. The length characteristic of the C$_{n}$  molecules is
according to $ZT\propto l^{2} \exp \left( -\xi l\right)$. The different length-scaling
behaviors between the thermal current conveyed by the electrons and the thermal current
conveyed by the phonons offer the key to the understanding of characteristic
temperature and the behavior of $ZT$. We believe that this study is a substantial
step towards the understanding of the thermoelectric properties in nanojunctions, and
we hope that this study will generate more experimental and theoretical explorations in
the properties of thermoelectric nanojunctions.

We are grateful to Prof. N. J. Tao for helpful discussions.The authors
thank MOE ATU, NCTS and NCHC for support under Grants NSC 97-2112-M-009-011-MY3,
097-2816-M-009-004 and 97-2120-M-009-005.


\begin{thebibliography}{99}
\bibitem{parametersAlatoms}  The aluminum (alkanethiol) junction consisted of an Al atomic chain (a C$_{4}$ molecule) sandwich between two Al (Au) electrodes that were modeled as electron jellium with $r_{s}\approx 2$ ($r_{s}\approx 3$). The
following parameters were used for the junctions: the effective
cross-section chosen was $A=23.1$~a.u.$^{2}$ ($21.4$~a.u.$^{2}$) for the aluminum (alkanetiol) junction. The spectral densities of electrodes were evaluated using the longitudinal and transverse
sound velocities for Al (Au), $v_{l}=6.35~(3.2)\times 10^{5}$~cm/s and $v_{t}=3.1~(1.2)\times
10^{5}$~cm/s, respectively. The Young's
modulus of aluminim (alkanethiol) wire, $Y\simeq 1.2\times 10^{13}$~dyne/cm$^{2}$ ($Y\simeq 2.3\times 10^{12}$~dyne/cm$^{2}$), was calculated with total energy calculations and was found to be almost independent of the length of the wires.
\end{thebibliography}
\end{document}